\documentclass[conference]{IEEEtran}
\IEEEoverridecommandlockouts
\usepackage{cite}
\usepackage{amsmath,amssymb,amsfonts}
\usepackage{algorithmic}
\usepackage{graphicx}
\usepackage{textcomp}
\usepackage{xcolor}
\usepackage{booktabs}
\usepackage{multirow}
\usepackage{arydshln}
\usepackage{makecell}
\usepackage{enumitem}
\usepackage{url}
\usepackage{hyperref}
\def\BibTeX{{\rm B\kern-.05em{\sc i\kern-.025em b}\kern-.08em
    T\kern-.1667em\lower.7ex\hbox{E}\kern-.125emX}}
\begin{document}

\title{FGSVQA: Frequency-Guided Short-form \\ Video Quality Assessment}

\author{Xinyi Wang, Angeliki Katsenou, Junxiao Shen, and David Bull\\
\textit{School of Computer Science, University of Bristol, Bristol BS1 8UB, UK}
}

\maketitle

\begin{abstract}
Short-form video poses new challenges to the quality assessment of user-generated content (UGC) due to its complex generation pipeline, rapid content variation, and mixed distortions. To address this challenge, we propose an end-to-end video quality assessment (VQA) framework that employs a dense visual encoder based on CLIP, and incorporates compression priors derived from the frequency domain to generate artifact- and structure-aware weight maps for feature aggregation. By explicitly decomposing artifact, structure, and original visual feature branches and adaptively fusing them over time through a learned gating module, the proposed method achieves accurate and efficient quality prediction. Experimental results show that our method achieves strong performance on short-form video datasets in terms of average rank and linear correlation (SRCC: 0.736, PLCC: 0.787), while maintaining efficient inference runtime. Code and additional results are available on \href{https://github.com/xinyiW915/FGSVQA}{GitHub}.
\end{abstract}


\section{Introduction}
Short-form (SF-) user-generated content (UGC) video has emerged as a mainstream media format, with rapidly growing popularity and accessibility among users~\cite{li2024ntire}. However, the complex content generation pipeline for short-form videos, including multiple pre-processing stages, poses new challenges for video quality assessment (VQA) of such content. On the one hand, the presence of diverse mixed distortions makes the analysis of quality degradation more challenging. However, rapid content variations make it difficult to perceive regions affected by quality degradation. These issues raise a question: Do objective VQA metrics designed for traditional UGC videos remain applicable to this new video format?

Existing datasets widely used for UGC quality assessment~\cite{nuutinen2016cvd2014, ghadiyaram2017capture, hosu2017konstanz, sinno2018large, wang2019youtube, ying2021patch, yu2023subjective, duan2025finevq} can be broadly divided into two categories: UGC videos captured under real-world conditions with authentic in-capture distortions~\cite{nuutinen2016cvd2014, ghadiyaram2017capture, hosu2017konstanz, sinno2018large}, and UGC videos collected from video-sharing platforms or streaming scenarios~\cite{wang2019youtube, ying2021patch, yu2023subjective, duan2025finevq}. These datasets contain videos with authentic distortions, providing a basis for VQA research. However, unlike conventional UGC videos, SF-UGC videos often span only a few seconds, are portrait-oriented, and typically feature rapid shot transitions and greater content variation. Moreover, the wide variety of creative modes on short-video platforms, such as special effects and kaleidoscopic content, together with complex post-processing techniques, including video enhancement and transcoding~\cite{lu2024kvq}, can further intensify quality fluctuations. Meanwhile, high dynamic range (HDR) content is now widely supported across various platforms and is becoming increasingly popular in SF-UGC videos. This further highlights the need to consider consumer viewing scenarios in which HDR content is converted to standard dynamic range (SDR)~\cite{wang2024sfv}.

Traditional full-reference VQA (FR-VQA) relies on the availability of pristine reference videos, and hence metrics such as PSNR, SSIM~\cite{wang2004image}, and VMAF~\cite{li2016toward} are not viable for UGC scenarios, where original reference videos are unavailable. In contrast, no-reference VQA (NR-VQA) does not require the original video, so is suitable for UGC quality assessment. Existing VQA methods have progressed from hand-crafted feature-based~\cite{korhonen2019two, tu2021ugc} approaches to deep learning-based models~\cite{tu2021rapique, li2022blindly}. Early methods relied on manually designed features and showed limited generalization to the diverse distortions in UGC. More recent studies have introduced 2D/3D CNNs~\cite{liu2018end, li2019quality, sun2022deep}, Transformers~\cite{you2021long, wu2023discovqa, wang2025frame}, and large multimodal models (LMMs)~\cite{wu2023towards, wu2023q, ge2024lmm}, improving quality prediction through temporal fusion~\cite{li2019quality, wu2023discovqa, yuan2023capturing}, multi-priors~\cite{li2022blindly, zhang2023md, liu2023ada}, and fragmentation~\cite{wu2023neighbourhood, wu2023exploring, wang2025diva, wang2026camp}. 

As SF-UGC often exhibits abrupt scene transitions, viewers focus mostly on the spatial information within these scenes~\cite{lu2024kvq, wu2023discovqa}. We propose a novel framework for SF-UGC videos that integrates dense visual features with frequency-domain priors. Since the discrete cosine transform (DCT) is widely used in video compression codecs, we use it to generate two spatial maps via frequency analysis: an \textit{artifact-aware} map that highlights regions susceptible to quality degradation, and a \textit{structure-aware} map that preserves the remaining spatial details. Using these maps as weights, the dense feature maps encoded by CLIP~\cite{radford2021learning} are aggregated through weighted pooling to produce three quality feature branches, which are then fused via a lightweight gating module for final quality prediction. 

The contributions of this paper are as follows:
\begin{enumerate}[leftmargin=*, itemsep=2pt, topsep=2pt, parsep=0pt]
    \item We propose a novel VQA model for short-form videos that employs frequency-guided weight maps to explicitly analyze different compression-related degradations, and aggregates visual representations from three quality-focused feature branches for the overall quality score.    
    \item Experimental results demonstrate that the proposed method achieves strong performance on SF-UGC datasets, and maintains high inference efficiency.
\end{enumerate}


\section{Proposed Method}
Given an input video, we uniformly sample $T$ frames over time. For each sampled frame $F_t$, a short temporal window is used to generate frequency-domain weight maps to emphasize regions more susceptible to compression distortion.

\begin{figure*}
    \centering
    \includegraphics[width=.8\linewidth]{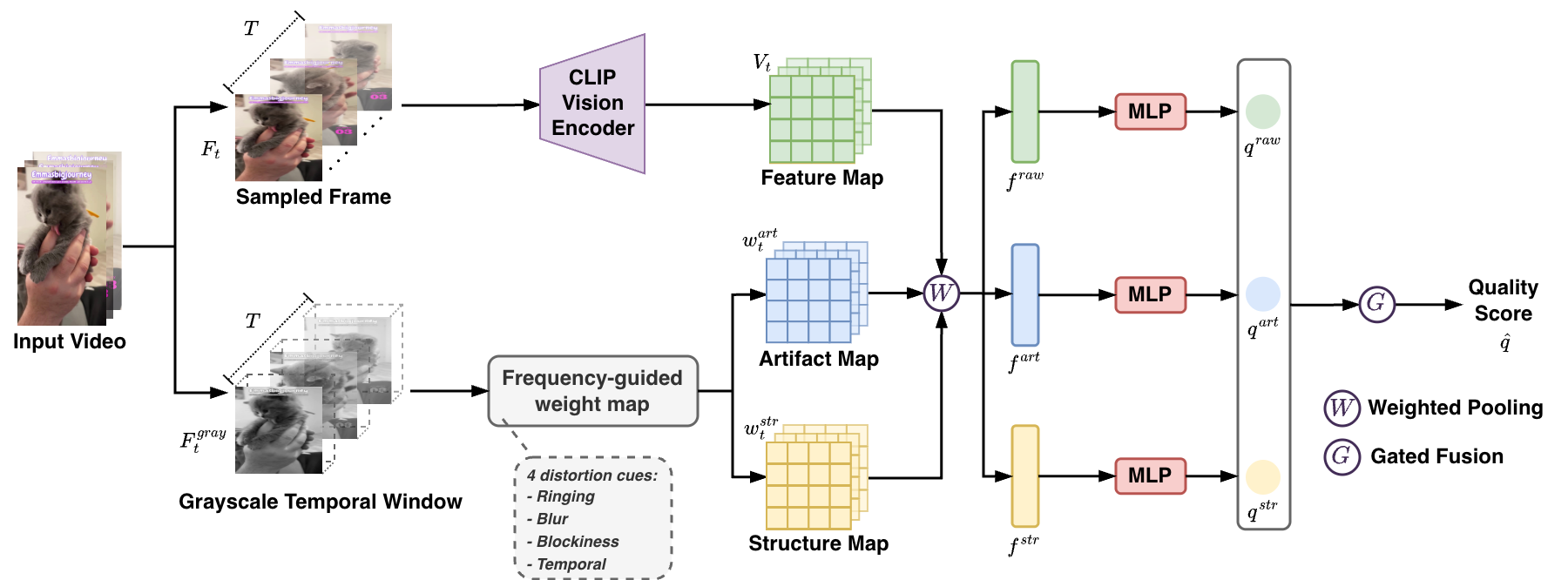}
    \caption{Overview of the proposed method with the two branches: the frequency-guided weight map and the CLIP vision encoder.}
    \vspace{-1.5em}
    \label{fig:placeholder}
\end{figure*}

\noindent
\textbf{Frequency-guided Weight Maps:}
We first convert frames within the temporal window to grayscale. Each grayscale frame $F_t^{\mathrm{gray}}$ is divided into non-overlapping $16 \times 16$ blocks, and a 2D DCT is applied to each block. Based on the DCT coefficients, we compute the normalized low-, mid-, and high-frequency energy ratios:
$r_c=\frac{E_c}{E}, c\in\{\text{low},\text{mid},\text{high}\}$
where $E_{\text{c}}$ is the frequency-band energy, $E$ is the total block energy.

Based on these block-wise spectral statistics, we focus on the dominant distortion artifacts:
\begin{enumerate}
    \item \textbf{Ringing} typically occurs near sharp edges and is associated with excessive mid- and high-frequency oscillations. We therefore extract a Sobel edge mask from $F_t^{\mathrm{gray}}$, compute the fraction of edge pixels within each block as a block-wise edge ratio, and combine it with the frequency energy $(r_{\text{mid}} + r_{\text{high}})$ while suppressing blocks with low edge presence.

    \item \textbf{Blur} mainly suppresses mid- and high-frequency details. We construct a blur prior from the complement of the weighted frequency energy $(0.5\,r_{\text{mid}} + r_{\text{high}})$ and modulate it with the block-wise Sobel gradient magnitude computed from $F_t^{\mathrm{gray}}$ to emphasize structured regions.

    \item \textbf{Blockiness} appears as unnatural intensity jumps across block boundaries. We measure horizontal and vertical intensity discontinuities across block boundaries, smooth the block-level boundary map with a Gaussian filter, and average it within each block to obtain a blockiness cue.

    \item \textbf{Temporal cue} captures quality fluctuations within the short temporal window. Within the window, we compute block-mean grids and apply the Fast Fourier Transform (FFT) along the temporal dimension. We then define motion as the ratio of non-DC temporal energy to DC energy, and flicker as the proportion of high-frequency temporal energy within the non-DC spectrum. These two components form the temporal distortion cue.
\end{enumerate}
These four cues are aggregated and normalized to form the artifact-aware map $w_t^\mathrm{art}$, while its complement serves as the structure-aware map $w_t^\mathrm{str}$ for relatively stable structural content. With frames resized to $224 \times 224$, the weight maps have height and width $H = W = 14$.

\noindent
\textbf{CLIP-based Quality Prediction Model:}
Each sampled frame $F_t$ is fed into a CLIP vision encoder to encode a visual feature map $V_t \in \mathbb{R}^{C \times H \times W}$, where $C$ is the channel dimension. Given the dense feature map $V_t$, we perform weighted spatial pooling using the frequency-guided map $W_t$, where $W_t \in \{w_t^{\mathrm{art}},\, w_t^{\mathrm{str}}\}$. The pooled feature is defined as:
\begin{equation}
z_t=\sum_{i,j}\tilde{W}_t(i,j)\,V_t(:,i,j),
\
\tilde{W}_t(i,j)=\frac{W_t(i,j)}{\sum_{m,n}W_t(m,n)},
\end{equation}
where $\tilde{W}_t$ denotes the normalized weights, and $i,j$ index the spatial locations on the dense map. By setting $W_t=w_t^{\mathrm{art}}$ and $W_t=w_t^{\mathrm{str}}$, we obtain two frame-level features, denoted as $z_t^{\mathrm{art}}$ and $z_t^{\mathrm{str}}$, respectively. In parallel, a raw visual feature is obtained by global average pooling on the same feature map:
\begin{equation}
z_t^{\mathrm{raw}}=\frac{1}{HW}\sum_{i,j}V_t(:,i,j).
\end{equation}
Over the $T$ sampled frames, features from each branch are temporally pooled as follows:
\begin{equation}
f^{b}=\frac{1}{T}\sum_{t=1}^{T} z_t^{b},
\qquad
b \in \{\mathrm{art},\mathrm{str},\mathrm{raw}\},
\end{equation}
where $T$ is the number of sampled frames.

\noindent
\textbf{Adaptive Fusion:}
Each aggregated feature is then fed into an independent three-layer MLP head to predict branch-wise quality scores, $q^{\text{art}}$, $q^{\text{str}}$, and $q^{\text{raw}}$. Each head consists of two fully connected hidden layers with ReLU activations and dropout, followed by a linear layer that outputs a scalar quality score. 

To adaptively adjust the contributions of the three branches, we employ a lightweight gated fusion. Given the mean values of the two weight maps and the mean absolute activation of the raw feature map, the gating module outputs softmax-normalized fusion weights $[\alpha,\beta,\gamma]$, where $\alpha,\beta,\gamma \in [0,1]$. The final quality score is computed as the weighted sum of the three branch scores:
\begin{equation}
\hat{q} = \alpha q_{\text{art}} + \beta q_{\text{str}} + \gamma q_{\text{raw}}.
\end{equation}

\section{Experiment}
\subsection{Evaluation setup}
\textbf{Datasets and Evaluation Metrics:}
We validated our proposed method on two publicly available SF-UGC datasets: KVQ~\cite{lu2024kvq} and the YouTube SFV+HDR dataset (YT-SFV)~\cite{wang2024sfv}. The YT-SFV dataset is the first publicly available dataset for SF-VQA, comprising 2,030 SDR videos and 2,000 SDR videos converted from HDR (HDR2SDR), spanning 10 content categories. KVQ is built on platform-processing pipelines to expand 600 user-uploaded short videos into 4,200 samples through pre-processing and transcoding. Datasets were randomly split into 80\%/20\% training and test sets (for KVQ, we followed the data split according to reference content used in NTIRE 2024 SF-UGC Challenge~\cite{li2024ntire}), with a validation set further split from the training set. The performance was evaluated using two widely adopted statistical metrics: Spearman Rank Correlation Coefficient (SRCC) and Pearson Linear Correlation Coefficient (PLCC).


\textbf{Implementation Details:}
Input videos were sampled into 16 frames, each with a 6-frame temporal window. The model was built on a CLIP ViT-B/16 visual encoder. Training used the AdamW optimizer for 35 epochs, with a batch size of 8, an initial learning rate of $1\times10^{-5}$, a learning rate of $5\times10^{-5}$ for unfrozen CLIP layers, and a weight decay of $1\times10^{-2}$. The CLIP encoder was frozen for the first 3 epochs, then the last four transformer blocks and the final layer normalization were unfrozen. The loss function combined the Smooth L1 and pairwise rank loss. The best checkpoint was saved based on validation SRCC, with early stopping at 6 patience epochs, and was used for testing. All experiments were run on an NVIDIA RTX 6000 Ada GPU.

\subsection{Performance Comparison}
\begin{table}[t]
\centering
\caption{Performance comparison of the evaluated NR-VQAs. \textbf{Bold} is the best result and \underline{underline} is second best.}
\label{tab:comparison}
\vspace{-0.5em}
\setlength{\tabcolsep}{3pt}
\renewcommand{\arraystretch}{1.15}
\begin{tabular}{@{}l|cc|cc|cc@{}}
\toprule
\multirow{2}{*}{Method}
& \multicolumn{2}{c|}{KVQ} & \multicolumn{2}{c|}{\makecell{YT-SFV\\ (SDR)}} & \multicolumn{2}{c}{\makecell{YT-SFV\\ (HDR2SDR)}} \\
\addlinespace[2pt]
& SRCC & PLCC & SRCC & PLCC & SRCC & PLCC \\
\midrule
FAST-VQA~\cite{wu2023neighbourhood}   & 0.832 & 0.834 & \textbf{0.789} & 0.789 & \textbf{0.543} & \underline{0.664} \\
FasterVQA~\cite{wu2023neighbourhood}  & N/A & N/A &0.748 & 0.753 & 0.493 & 0.585 \\
DOVER~\cite{wu2023exploring}          & 0.833 & 0.837 & 0.750 & \underline{0.793} & 0.496 & 0.618 \\
KSVQE~\cite{lu2024kvq}                & \underline{0.867} & \underline{0.869} & N/A & N/A & N/A & N/A \\
\midrule
\textbf{FGSVQA} & \textbf{0.877} & \textbf{0.878} & \underline{0.788} & \textbf{0.818} & \textbf{0.543} & \textbf{0.666} \\
\bottomrule
\end{tabular}
\vspace{-1.9em}
\end{table}
We compared our model with state-of-the-art (SOTA) methods, including NR-VQA for general UGC~\cite{wu2023neighbourhood, wu2023exploring} and KSVQE~\cite{lu2024kvq} for SF-UGC. As shown in Table~\ref{tab:comparison}, FGSVQA achieves the best overall performance on SF-VQA, especially on KVQ with complex effects and kaleidoscopic content, attaining the highest SRCC and PLCC of 0.877 and 0.878. For SFV (SDR), Fast-VQA achieves a high SRCC of 0.789, indicating that general NR-VQA models can transfer to SF-scenarios, though they may miss unique quality features. FGSVQA also remains competitive, achieving the highest PLCC of 0.818. For YT-SFV (HDR2SDR), FGSVQA obtains the best PLCC of 0.666, while SRCC of 0.543 matches Fast-VQA. 

Table~\ref{tab:cross_dataset_eval} presents the cross-dataset evaluation of FGSVQA. The upper part reports direct transfer results of models trained on different SF-datasets, and the lower part reports the results after fine-tuning the corresponding source-trained checkpoints on each target dataset. For example, when trained on KVQ, FGSVQA achieves 0.755/0.807 SRCC/PLCC on YT-SFV (SDR), but only 0.476/0.589 on YT-SFV (HDR2SDR). Fine-tuning the same KVQ-trained checkpoint consistently improves performance to 0.829/0.868 and 0.641/0.723, respectively. All models show relatively weak correlations on HDR2SDR, suggesting that HDR-converted videos are harder to assess due to the skewed quality distribution (90\% of quality score $>$ 4.0) and the stronger color sensitivity in HDR content.

\begin{table}[t]
\centering
\caption{Cross-dataset evaluation of FGSVQA.}
\label{tab:cross_dataset_eval}
\vspace{-0.5em}
\setlength{\tabcolsep}{2pt}
\renewcommand{\arraystretch}{1.15}
\begin{tabular}{@{}l|cc|cc|cc@{}}
\toprule
\multicolumn{1}{r|}{Train on: } & \multicolumn{2}{c|}{KVQ} & \multicolumn{2}{c|}{\makecell{YT-SFV\\ (SDR)}} & \multicolumn{2}{c}{\makecell{YT-SFV\\ (HDR2SDR)}} \\
\addlinespace[2pt]
Test on: & SRCC & PLCC & SRCC & PLCC & SRCC & PLCC \\
\midrule
KVQ                     & -- & -- & 0.734 & 0.745 & 0.598 & 0.569\\
YT-SFV (SDR)       & 0.755 & 0.807 & -- & -- & 0.617 & 0.680\\
YT-SFV (HDR2SDR)   & 0.476 & 0.589 & 0.545 & 0.696 & -- & -- \\

\midrule
Finetune on: \\
KVQ                     & -- & -- & 0.886 & 0.888 & 0.874 & 0.879 \\
YT-SFV (SDR)       & 0.829 & 0.868 & -- & -- & 0.818 & 0.856 \\
YT-SFV (HDR2SDR)   & 0.641 & 0.723 & 0.659 & 0.801 & -- & -- \\
\bottomrule
\end{tabular}
\vspace{-1em}
\end{table}

Finally, we measured the runtime of processing the same video at different spatial resolutions. For a fair comparison, we trained FGSVQA on LSVQ~\cite{ying2021patch}, as did the other general UGC quality metrics, and evaluated all models on a sample SDR video from YT-SFV. The results reported in Table~\ref{tab:complexity} show that FGSVQA maintains low runtime and high inference efficiency across resolutions, taking 0.31 seconds at 540P and 2.1 seconds at 2160P while still providing reliable quality prediction.

\begin{table}[t]
\centering
\caption{GPU runtime comparison (averaged over 10 runs) across different resolutions on "SDR\_Animal\_5ngj.mp4".}
\label{tab:complexity}
\vspace{-0.5em}
\setlength{\tabcolsep}{4pt}
\renewcommand{\arraystretch}{1.15}
\begin{tabular}{@{}l|c|c|c|c|c@{}}
\toprule
& \multicolumn{4}{c|}{Time(s)} & Ground truth: 4.308\\ \addlinespace[2pt]
Method: & 540P & 720P & 1080P & 2160P & Predicted Score \\
\midrule
Fast-VQA~\cite{wu2023neighbourhood}    & 0.599 & 0.673 & 0.909 & 2.217 & 3.319\\
FasterVQA~\cite{wu2023neighbourhood}   & \underline{0.489} & \underline{0.547} & \textbf{0.696} & \textbf{1.343} & 3.556\\
DOVER~\cite{wu2023exploring}           & 0.920 & 1.022 & 1.293 & 2.783 & \underline{3.814}\\
\midrule
FGSVQA                                  & \textbf{0.313} & \textbf{0.405} & \underline{0.697} & \underline{2.137} & \textbf{3.878}\\
\bottomrule
\end{tabular}
\vspace{-2em}
\end{table}

\section{Conclusion}
We have proposed a new NR-VQA model for SF-UGC based on a CLIP encoder enhanced with frequency-guided weight maps. By decomposing feature maps into three quality branches and integrating features through gated fusion, our model effectively captures the spatio-temporal distortions caused by compression. Experiments on two SF-UGC datasets showed that the proposed method outperforms SOTA NR-VQA UGC methods on average, achieving an SRCC of 0.736 and a PLCC of 0.787, while delivering the fastest inference time at low resolutions. These results also underscore the need for further development tailored to SF-UGC features. Future work should thus focus on studying the data manifold of SF-UGC to further improve metric design and performance.

\section*{Acknowledgment}
This work was funded by the UKRI MyWorld Strength in Places Programme (SIPF00006/1).

\bibliographystyle{IEEEtran}
\bibliography{refs}
\end{document}